\documentclass[twocolumn,showpacs,preprintnumbers,amsmath,amssymb,floatfix,prb]{revtex4}

\usepackage{epsfig,psfrag}
\usepackage{dcolumn}
\usepackage{bm}
\usepackage{graphicx}
\usepackage{color}
\newcommand{\gz}{\mathbb{Z}} 

\begin{document}

\title{Spin-orbit coupling and spectral function of interacting electrons in carbon nanotubes}
\author{Andreas Schulz,$^1$ Alessandro De Martino,$^{2}$ and 
Reinhold Egger$^1$ }    
\affiliation{${}^1$~Institut f\"ur Theoretische Physik, 
Heinrich-Heine-Universit\"at,
D-40225  D\"usseldorf, Germany \\
${}^2$~Institut f\"ur Theoretische Physik, Universit\"at zu K\"oln,
D-50937 K\"oln, Germany}

\date{\today}

\begin{abstract}

The electronic spin-orbit coupling in carbon nanotubes is strongly enhanced by
the curvature of the tube surface and has important effects on the 
single-particle spectrum. Here, we include the full spin-orbit interaction in 
the formulation of  the effective low-energy theory for
interacting electrons in metallic single-wall carbon nanotubes and study 
its consequences. The resulting theory is a four-channel Luttinger liquid, where 
spin and charge modes are mixed. We show that the analytic structure
of the spectral function is strongly affected by this mixing, which can
provide an experimental signature of the spin-orbit interaction.

\end{abstract}

\maketitle

Spin-orbit interaction (SOI) effects\cite{winkler} are of great interest in the 
field of spintronics, and their detailed understanding is both of fundamental 
and of technological interest, e.g., for the coherent
manipulation of spin qubits.\cite{spinqubit}
In single-wall carbon nanotubes (SWNTs) the SOI 
arises predominantly from the interplay of atomic SO coupling 
and curvature-induced hybridization, and its effect on the electronic 
bandstructure has recently been clarified.\cite{andoso,prlale,murphy,
chico,rashbaale,paco,zhou,jeong,saito}
Contrary to other carbon-based materials, as, e.g., flat graphene, 
where SOI is very weak (of the order of few $\mu$eV), in 
SWNTs it can reach fractions of meV and has important consequences, 
previously overlooked. 
New experiments \cite{mceuen,churchill} on ultraclean SWNTs, 
made possible by advances in the fabrication technology, have indeed
observed modifications of the electronic spectrum due to
SOI. These results confirm the theoretical expectations, and motivate 
renewed interest on SOI in nanotubes. 
So far on the theory side the main focus has been on nanotube quantum dots,\cite{cntsodots}
where the SOI manifests itself in spectral features. Here we study 
long SWNTs, where long-ranged interactions can induce 
non-Fermi liquid electronic phases.\cite{ando}  In particular, 
without SOI, Luttinger liquid (LL) behavior\cite{gogolin} has been
predicted for metallic SWNTs.\cite{egger97}
Experimental evidence for this strongly correlated phase has been
reported using quantum transport\cite{bockrath} and photoemission 
spectroscopy.\cite{ishii}  

The question therefore arises how the LL theory of SWNTs\cite{egger97}  
is modified when the SOI is taken into account, and what are 
its observable consequences.
This question is addressed and answered in detail below.
Our main results are as follows.  (i) The low-energy
theory of metallic SWNTs still describes a Luttinger liquid.
However, the decoupled plasmon modes do not correspond to spin and charge anymore.
\textit{Spin-charge separation} in the usual sense\cite{gogolin} is 
therefore \textit{broken} by the SOI.
This effect can be traced back to a term in the SO Hamiltonian diagonal in sublattice space 
(see Eq. \ref{h0}), which was previously overlooked.
(ii)  We discuss in detail the \textit{spectral function},
a quantity that can directly be probed experimentally for SWNTs.\cite{ishii}  
We show how the mixing of spin and charge modes due to SOI affects 
its analytic structure and modifies it from the established  spinful LL 
behavior\cite{gogolin,meden}.  The predicted deviations
are small but should be observable. 
(iii) The tunneling density of states, and hence most typical
quantum transport observables, is only weakly affected by the SOI. 
This may explain why the SOI in SWNTs has long been overlooked. 
(iv) We shall clarify the similarities and the differences 
of the present SWNT theory to the LL description of 1D interacting 
semiconductor wires  with Rashba SOI.\cite{moroz,epl,governale,hausler,gritsev,oleg1,rkky}

To start, let us address the bandstructure of a 
nominally metallic $(n,m)$ SWNT, where  
$2n+m\in 3\gz$. The chiral angle\cite{ando}
is $\theta=\tan^{-1}[\sqrt{3}m/(2n+m)]$,
and the tube radius is $R [{\rm nm}] \simeq 0.0391 \sqrt{n^2+nm+m^2}$. 
We employ the effective SO Hamiltonian for $\pi$ electrons
derived in Ref.~\onlinecite{saito} in the ${\bm k}\cdot {\bm p}$ scheme.  
This model is in semi-quantitative accordance with available
experimental Coulomb blockade spectroscopy data,\cite{mceuen}
and summarizes earlier theoretical work.  In particular,
it includes the recently discovered ``diagonal'' contribution $E_{\rm SO}$, 
which is of crucial importance in our analysis (see below).  
Within this framework, the single-particle Hamiltonian
$H_0({\bm k})$ for wavevector ${\bm k}= (k,k_\perp)$ relative to the 
respective $K$ point
is a $2\times 2$ matrix in sublattice space corresponding
to the two basis atoms of the honeycomb lattice. This 
separately applies to both $K$ points $\alpha=\pm$ and both 
spin directions $\sigma=\pm$, where the spin quantization axis is along the 
tube axis. To leading order in the SOI, the 
spin label $\sigma$ is still a good quantum number.\cite{andoso,prlale,paco}
Periodic boundary conditions around the SWNT circumference 
imply a quantization of transverse momentum, $k_\perp R = n_0 \in \gz$.
We assume a Fermi energy $E_F>0$ but
sufficiently small to justify that only the $n_0=0$ band has to be 
retained.  All other bands are then separated by an energy gap 
$\approx\hbar v_F/R\approx 1$~eV, where
 $v_F\approx 8\times 10^5$~m/s. 
Then, $H_0(k)$  is given by\cite{saito}
\begin{equation}\label{h0}
 \left( \begin{array}{cc} \alpha\sigma E_{\rm SO} & 
-\alpha\hbar v_F [\phi_\perp +
i(k+\alpha\phi_\parallel )] \\
-\alpha\hbar v_F[\phi_\perp-i(k+\alpha\phi_\parallel)] 
& \alpha\sigma E_{\rm SO} \end{array}
\right).
\end{equation}
This form neglects trigonal warping corrections,\cite{ando} which 
cause only tiny changes in the low-energy physics
but would complicate our analysis substantially.
Using the parameter estimates of Ref.~\onlinecite{saito},
the diagonal term is
\begin{equation}\label{estimate}
E_{\rm SO}[\textrm{meV}] \simeq -\frac{0.135 \cos(3\theta) }{R[\textrm{nm}]} .
\end{equation}
Writing $\phi_\perp =\phi_{\perp,{\rm SO}} + \phi_{\perp,{\rm cur}}$,
the SOI corresponds to a spin-dependent  
shift of the transverse momentum,\cite{paco,mceuen,saito}
\begin{equation}
\phi_{\perp,{\rm SO}} [\textrm{nm}^{-1}]
\simeq \alpha \sigma \frac{2.7 \times 10^{-4}}{R[\textrm{nm}]} , 
\end{equation}
while curvature effects\cite{saito,ando} give 
\begin{eqnarray*}
\phi_{\perp,{\rm cur}} [\textrm{nm}^{-1}]
&\simeq & \frac{0.011 \cos(3\theta)}{(R[\textrm{nm}])^2} , \\
\phi_{\parallel} [\textrm{nm}^{-1}]
&\simeq & \frac{0.045 \sin(3\theta)}{(R[\textrm{nm}])^2} .
\end{eqnarray*}
We remark that the Hamiltonian (\ref{h0}) contains the two leading
effects of curvature-induced SOI, namely the diagonal contribution 
$E_{\rm SO}$ and the Rashba-type SOI encoded by
$\phi_{\perp,{\rm SO}}$.  Subleading terms, e.g., the 
``intrinsic'' SOI,\cite{paco} are much smaller and  not
 taken into account here.
The dispersion relation obtained from Eq.~(\ref{h0}) is
\begin{equation}\label{ee}
E_\pm^{(\alpha,\sigma)} (k) = \alpha\sigma E_{\rm SO}
\pm \hbar v_F\sqrt{\phi_\perp^2+ (k+\alpha\phi_\parallel)^2} ,
\end{equation}
where the Kramers degeneracy is reflected in $E_{\pm}^{(\alpha,\sigma)}(k)
=E^{(-\alpha,-\sigma)}_{\pm}(-k)$. 
Since $E_F>0$, only the conduction bands (positive sign) are kept, and
the Fermi momenta $k^{(F)}_{r\alpha\sigma}$ for right- and left-movers
($r=R/L=\pm$) follow from
$E^{(\alpha,\sigma)}_+ \left(k^{(F)}_{r\alpha\sigma} \right)=E_F$, 
$k^{(F)}_{r\alpha\sigma}\approx r(E_F-\alpha\sigma E_{\rm SO})/\hbar v_F-\alpha \phi_\parallel$.
We linearize the dispersion relation around
the Fermi points, always assuming that $E_F$ is sufficiently
far away from the band bottom.  The 1D Fermi velocities
$v_{\alpha,\sigma}= \hbar^{-1}
\partial_{k} E^{(\alpha,\sigma)}_+\left(k=k^{(F)}_{
+,\alpha\sigma}\right)$  take only two different values,
$v_A \equiv v_{-,\uparrow}=v_{+,\downarrow}$ and
$v_B \equiv v_{+,\uparrow}=v_{-,\downarrow}$.
We mention in passing that
$R/L$ movers have pairwise identical velocities only
in the absence of trigonal warping and
orbital magnetic fields\cite{tsvelik} or transverse fields,\cite{vishvesh}
as assumed here.
It is convenient to introduce the mean velocity
$v=(v_A+v_B)/2$ and the dimensionless difference $\delta = (v_A-v_B)/(2v)$.
After some algebra, Eq.~(\ref{ee}) together with the 
parameter estimates above yields 
\begin{eqnarray}\nonumber
\frac{v}{v_F} &\simeq& 1-\frac{0.01 (R [{\rm nm}])^2 + 17 \cos^2(3\theta)} 
{ (E_F[{\rm meV}])^2 \ (R [{\rm nm}])^4} ,\\ \label{delta} 
\delta &\simeq& \frac{0.83 \cos (3\theta) }{(E_F[{\rm meV}])^2 \
 (R [{\rm nm}])^3}  .
\end{eqnarray}
The renormalization of $v$ away from $v_F$ goes always downwards,
but the quantitative shift is small.
The asymmetry parameter $\delta$ effectively parametrizes the SOI strength
and is more important in what follows.
For fixed $E_F$ and $R$, it is maximal for $\theta=0$ (zig-zag tube)
and vanishes for $\theta=\pi/6$ (armchair tube). 
Moreover, $\delta$ increases for smaller tube radius, but the continuum
description underlying our approach eventually breaks down for $R\alt
0.4$~nm.  Since $E_F$ should at the same time be sufficiently far 
above the band bottom in Eq.~(\ref{ee}), 
in practice this leads to rather small values, $\delta\alt 0.05$. 
This is a rather conservative estimate, though, based on the parameter
values of Ref. \onlinecite{saito} and larger values could be obtained 
if one uses different estimates. Nonetheless, we show below that observable 
consequences do arise.

The theory is then equivalently formulated using
 Abelian bosonization,\cite{gogolin} which allows for the 
nonperturbative inclusion of interactions. 
We employ the boson fields $\phi_\alpha(x)$ with
 $\alpha=c+,c-,s+,s-$, representing the 
total and relative charge and spin density modes,\cite{egger97} and 
their conjugate momentum fields $\Pi_\alpha(x)=- \partial_x\theta_{\alpha}$,
where $\theta_\alpha$ are the dual fields.
Those fields are conveniently combined into the vectors 
$\Phi_I(x)=(\phi_{c+},\theta_{c+},\phi_{s-},\theta_{s-})^T$
and $\Phi_0(x)= (\phi_{c-},\theta_{c-},\phi_{s+},\theta_{s+})^T$.
The important electron-electron forward scattering\cite{foot2}
 effects are parametrized by the standard LL parameter $K\equiv K_{c+}$, 
where $K=1$ for noninteracting electrons but $K\approx 0.2\ldots 0.4$  
for SWNTs deposited on insulating substrates
(or for suspended SWNTs) due to the long-ranged Coulomb 
interaction.\cite{egger97,bockrath,ishii}  
The low-energy Hamiltonian of a spin-orbit-coupled interacting
metallic SWNT then reads
\begin{equation}\label{ham}
H =\frac{\hbar v}{2}\int dx \left (\begin{array}{c}\partial_x\Phi_I
\\ \partial_x\Phi_0\end{array}\right)^T 
\left(\begin{array}{cc} {\bm h}(K) & 0 \\ 0 & {\bm h}(1) \end{array}
\right)
\left(\begin{array}{c}  \partial_x\Phi_I\\ \partial_x\Phi_0\end{array}
\right)
\end{equation}
with the $K$-dependent matrix
\[
{\bm h}(K) =
\left(\begin{array}{cccc} \frac{1}{K^2} & 0 & \delta & 0 \\ 
0 & 1 & 0 & \delta\\ \delta & 0 & 1 & 0\\
0 & \delta & 0 & 1 \end{array} \right).
\]
The above representation shows that SOI ($\delta\ne 0$)
breaks spin SU(2) symmetry.  Notably, the modes $\Phi_I$ and $\Phi_0$ decouple, 
and interactions ($K\ne 1$) only affect the $\Phi_I$ sector.  
In each sector, the Hamiltonian is then formally identical to the one for a 
semiconductor wire with Rashba SOI in the
 absence of backscattering.\cite{rkky}  
 We consider a very long SWNT and ignore finite-length effects, i.e. the
 zero modes contributions to the Hamiltonian (\ref{ham}).

Equation (\ref{ham}) can be diagonalized by the linear 
transformation\cite{kimura} $\Phi_I={\bm V}_I\Phi_a$ and 
$\Phi_0={\bm V}_0 \Phi_b$, with the $4\times 4$ matrix
\begin{equation}\label{trafo}
{\bm V}_I = \left( \begin{array}{cccc}
\cos\eta & 0 & -\frac{\sin\eta}{y} & 0\\
0& \cos\eta & 0 & -y  \sin\eta\\
y\sin\eta & 0 & \cos\eta & 0\\
0 & \frac{\sin\eta}{y}&0 &\cos\eta
\end{array}\right),
\end{equation}
where 
\begin{equation}\label{ydef}
y=\sqrt{(1+K^{-2})/2},\quad \tan(2\eta)
 =\frac{2\delta y}{y^2-1}.
\end{equation}
 ${\bm V}_0$ is as in Eq.~(\ref{trafo}) with $K=1$, 
 i.e., $y=1$ and $\eta=\pi/4$.
In terms of the new vectors $\Phi_{\rho}
= (\phi_{+,\rho},\theta_{+,\rho},
\phi_{-,\rho},\theta_{-,\rho})^T$ with mutually dual boson fields 
$\phi_{j\rho}$ and $\theta_{j\rho}$ for each set $(j=\pm,\rho=a/b)$,
the diagonalized Hamiltonian is seen to describe a 
four-channel Luttinger liquid,
\begin{equation} \label{ham2}
H =\sum_{j,\rho} \frac{\hbar v_{j\rho}}{2} 
\int dx \left(\frac{1}{K_{j\rho}} 
(\partial_{x} \phi_{j\rho})^2 + K_{j\rho} (\partial_x \theta_{j\rho})^2
\right).
\end{equation}
The interacting sector corresponds to $\rho=a$, where
the effective LL parameters $K_{\pm,a}$ and
the plasmon velocities $v_{\pm,a}$  are 
\begin{eqnarray}
K_{\pm,a} &=& y^{\mp 1}\sqrt{\frac{3+K^{-2}\pm \Delta}{3K^{-2}+1\pm
\Delta}},\\ \nonumber \frac{v_{\pm,a}}{v}
 &=& \sqrt{y^2+\delta^2\pm \Delta/2},\\
\nonumber
\Delta &=& \sqrt{(K^{-2}-1)^2 +(4\delta y)^2}
\end{eqnarray}
with $y$ in Eq.~(\ref{ydef}).  For $\rho=b$, the noninteracting values 
apply, $K_{\pm,b}=1$ and $v_{\pm,b}=v(1\pm \delta)$.
Note that the above expressions recover the LL theory
for $\delta=0$,\cite{egger97}
 where $v_{j\rho}=v_F/K_{j\rho}$ with $K_{j\rho}=1$ except for $K_{+,a}=K$.

Within the framework of the LL Hamiltonian (\ref{ham2}),
using the bosonized form of the electron field operator\cite{gogolin,egger97}
$\Psi_{r\alpha\sigma}(x,t)$ and the transformation (\ref{trafo}), 
it is possible to obtain exact results for all observables of interest.
In particular, arbitrary correlation functions of exponentials of
the boson fields can be calculated.  As an important application,
we discuss here the \textit{spectral function}
for an $r=R/L$ moving electron
with spin $\sigma$ near the $K$ point $\alpha=\pm$, which is defined as 
\begin{equation}\label{spectralfunction}
A_{r\alpha\sigma}(q,\omega) = -\frac{1}{\pi}\ 
{\rm Im}\ G^{\rm ret}_{r\alpha\sigma}(q,\omega),
\end{equation}
with the Fourier transform of the single-particle retarded 
Green's function ($\Theta$ is the Heaviside function),
\[
G^{\rm ret}_{r\alpha\sigma}(x,t) =-i\Theta(t)\left[\left\langle
\Psi^{}_{r\alpha\sigma}(x,t)\Psi^\dagger_{r\alpha\sigma}
(0,0)\right\rangle+ {\rm c.c.}\right],
\]
and the momentum $q$
is measured with respect to the relative Fermi momentum $k^{(F)}_{r\alpha\sigma}$. 

After some algebra, Eq.~(\ref{spectralfunction}) follows in closed 
form, which we specify in the zero-temperature limit now.
With the short-distance cutoff (lattice spacing) $a_0\approx 0.246$~nm, we find
\begin{align}\label{specf}
&A_{r\alpha\sigma}(q,\omega) \propto  \int_{-\infty}^\infty
dx \int_{-\infty}^\infty dt \ e^{-i[q x - \omega t]}\\ \nonumber 
&\times \Biggl [\prod_{j,\rho} \prod_{\mu=\pm} 
\left( 1+i \frac{ v_{j\rho} t +\mu r x}{a_0}
\right)^{-\Gamma_{j,\rho;\mu}^{(\alpha\sigma)}} 
+  (x,t)\to (-x,-t)\Biggr]
\end{align}
where the exponents for $j=\pm$ and $\mu=\pm$ are
given by [see also Eq.~(\ref{ydef})]
\begin{eqnarray}\label{exponents}
\Gamma_{j,a;\mu}^{(\alpha\sigma)} &=& \frac{1}{16} \Bigl [ \cos(\eta) 
\left(K_{j,a}^{1/2} -\mu K_{j,a}^{-1/2} \right) 
\\ \nonumber &+& \alpha \sigma j \sin(\eta) \left( y^{j} K_{j,a}^{1/2}
-\mu y^{-j} K^{-1/2}_{j,a} \right) \Bigr]^2,
\\ \nonumber
\Gamma_{j,b;\mu}^{(\alpha\sigma)}
&=& \frac12 \delta_{j,\alpha\sigma} \delta_{\mu,-}.
\end{eqnarray}
The remaining Fourier integrals are difficult to perform.  We here
follow Ref.~\onlinecite{gogolin} and focus on the analytic structure
of the spectral function,
which can be obtained by the power counting technique and Jordan's lemma.   
Up to an overall prefactor, the spectral function 
exhibits power-law singularities close to the
lines $\omega= \pm v_{j\rho} q$.  These singularities are 
captured by the approximate form
\begin{eqnarray} \label{singular}
A_{r\alpha\sigma}(q,\omega) &\approx &
\left(\prod_{j,\mu}\left |\omega +\mu r v_{j,a} q
\right|^{\Gamma^{(\alpha\sigma)}-1-
\Gamma_{j,a;\mu}^{(\alpha\sigma)}} \right) \\ \nonumber 
&\times& \left |
\omega-r(1+\alpha \sigma\delta) v q\right |^{\Gamma^{(\alpha\sigma)}-3/2}
 \\ \nonumber
&\times& \left [ \Theta(\omega-r\bar v q)+\Theta( -\omega - r v_{-a} q) \right],
\end{eqnarray}
where $\bar v = {\rm min}[v_{-,a},(1+\alpha\sigma \delta) v]$ and
\begin{equation}\label{Gammadd}
\Gamma^{(\alpha\sigma)} = \sum_{j\rho\mu} \Gamma^{(\alpha\sigma)}_{j\rho\mu}.
\end{equation}
We stress that Eq.~(\ref{singular}) is asymptotically exact:
it has the same analytic structure and the same exponents  of the power
laws at the singular lines $\omega=\pm v_{j\rho} q$ as the exact spectral
function. Away from the singularities, however, it  only serves illustrative purposes.

\begin{figure}
\scalebox{0.35}{\includegraphics{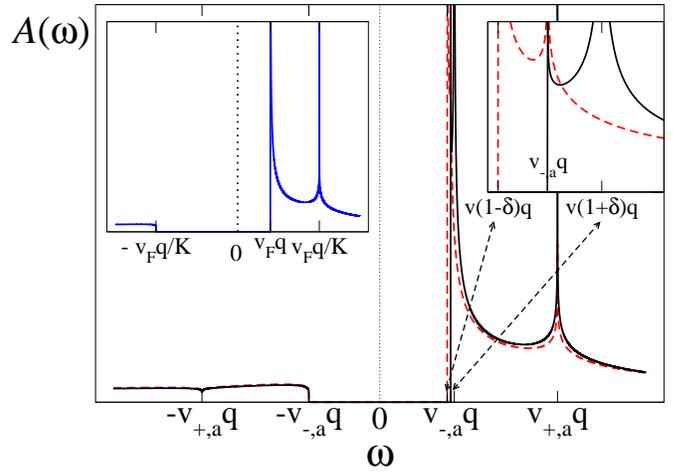}} 
\caption{ \label{fig1}  (Color online) 
Spectral function (\ref{singular}) for a right-mover in an interacting 
SWNT with LL parameter $K=0.4$ and SOI parameter $\delta=0.05$,
 shown in arbitrary units as function of $\omega$ for
given wavevector $q>0$. 
The black solid curve is for $\alpha\sigma=+1$, while the
red dashed curve is for $\alpha\sigma=-$.
Note that $A_{R\alpha\sigma}(q,\omega)=0$ for $-v_{-,a}q<\omega<\bar v q$. 
Right inset:  Magnified view around $\omega\approx v_{-,a}q$. 
Left inset:  Same as main panel but without SOI ($\delta=0$).  
Shifts of the positions of the singularities due to the shifts of Fermi momenta 
are not included in the figure since each spectral function  $A_{r\alpha\sigma}(q,\omega)$ is 
evaluated at momentum $q$ relative to the {\em respective} Fermi momentum. 
} 
\end{figure}

The spectral function (\ref{singular}) is depicted 
in the main panel of Fig.~\ref{fig1} for fixed 
wavevector $q>0$ as a function of frequency $\omega$, 
taking $K=0.4$ and $\delta=0.05$.
Compared to the well-known spectral function in the absence of 
SOI ($\delta=0$), see left inset of Fig.~\ref{fig1} and 
Refs.~\onlinecite{gogolin,meden}, additional structure 
can be observed for $\delta\ne 0$.  First, the singular
feature around $\omega=v_{-,a}q$ splits into two
different power-law singularities when $\delta\ne 0$, see
the right inset of Fig.~\ref{fig1} for a magnified view.  For large 
$q$, the corresponding frequency differences are in the meV regime and
can be resolved even for the rather small $\delta$ expected here.
Second,  for $-v_{+,a}q<\omega<
-v_{-,a}q$, the spectral function is finite (albeit small) when
$\delta\ne 0$.  Note that for $\delta=0$, the respective velocities 
are $v_{+,a}=v_F/K$ and $v_{-,a}=v_F$, implying a large frequency
window where this effect may take place.  
These predictions for the spectral function could be detected by 
photoemission spectroscopy. 

Many standard quantum transport properties, however, will hardly
show an effect due to the SOI, which may explain why effects of SOI in SWNTs
have been so long overlooked. For instance, the tunneling density of states 
averaged over $(r,\alpha,\sigma)$
exhibits power-law scaling with $\omega$ for low frequencies,
$\nu(\omega)\propto \omega^{\gamma-1}$. 
The exponent $\gamma$ is the smaller of the quantities
$\Gamma^{(\pm)}$ in Eq.~(\ref{Gammadd}).
This exponent is analytic in $\delta$, and the smallness of 
$\delta$ then implies that the tunneling density of states
in SWNTs will be very close to the one in the absence of SOI.
Let us also briefly comment on the relation of our results to the LL
theory for semiconductor quantum wires with Rashba
SOI.\cite{moroz,epl,governale,hausler,gritsev,oleg1,rkky}
The ``interacting'' sector $\rho=a$ in Eq.~(\ref{ham2}) coincides
with the semiconductor theory when electron-electron backscattering
can be neglected.  The additional presence of the 
``noninteracting'' sector $\rho=b$, however, causes additional
structure in the spectral function.  Moreover, while  
backscattering in semiconductor wires is likely an 
irrelevant perturbation in the renormalization group sense,\cite{rkky}
it nonetheless causes a renormalization of the LL parameters
and the plasmon velocities.  Such renormalization effects are
negligible in SWNTs.

To conclude, we have studied SOI effects
on the effective low-energy theory of interacting metallic 
SWNTs. We have shown that a four-channel Luttinger liquid theory remains
applicable, but compared to the previous formulation without SOI,\cite{egger97} 
all four channels are now characterized by different Luttinger liquid parameters 
and plasmon velocities, reflecting the broken spin $SU(2)$ symmetry. 
The coupling of spin and charge modes leads then to observable
modifications in the spectral function, which provide an experimental 
signature of SOI. 
This work was supported by the SFB TR 12 of the DFG.


\begin{thebibliography}{99}

\bibitem{winkler}
R. Winkler, \textit{Spin-Orbit Coupling Effects in Two-Dimensional
Electron and Hole Systems} (Springer, Berlin, 2003).

\bibitem{spinqubit}
C. Flindt, A.S. S\o rensen, and K. Flensberg,
Phys. Rev. Lett. {\bf 97}, 240501 (2006);
D.V. Bulaev, B. Trauzettel, and D. Loss, Phys. Rev. B {\bf 77}, 235301 (2008).

\bibitem{andoso}
T. Ando, J. Phys. Soc. Jpn. {\bf 69}, 1757 (2000).

\bibitem{prlale}
A. De Martino, R. Egger, K. Hallberg, and C.A. Balseiro, Phys. Rev. Lett.
{\bf 88}, 206402 (2002).

\bibitem{murphy}
A. De Martino, R. Egger, F. Murphy-Armando, K. Hallberg,
J. Phys. Cond. Matt. {\bf 16}, S1437 (2004).

\bibitem{chico}
L. Chico, M.P. Lopez-Sancho, and M.C. Munoz, Phys. Rev. Lett. {\bf 93}, 
176402 (2004); Phys. Rev. B {\bf 79}, 235423 (2009).

\bibitem{rashbaale}
A. De Martino and R. Egger, J. Phys. Cond. Matt. {\bf 17}, 5523 (2005).

\bibitem{paco}
D. Huertas-Hernando, F. Guinea, and A. Brataas, Phys. Rev. B {\bf 74},
155426 (2006).

\bibitem{zhou}
J. Zhou, Q. Liang, and J. Dong, Phys. Rev. B {\bf 79}, 195427 (2009).

\bibitem{jeong}
J.S. Jeong and H.W. Lee, Phys. Rev. B {\bf 80}, 075409 (2009).

\bibitem{saito}
W. Izumida, K. Sato, and R. Saito, J. Phys. Soc. Jpn. {\bf 78}, 074707 (2009).

\bibitem{mceuen}
F. Kuemmeth, S. Ilani, D.C. Ralph, and P.L. McEuen, Nature {\bf 452}, 448 
(2008).

\bibitem{churchill}
H.O.H. Churchill, F. Kuemmeth, J.W. Harlow, A. J. Bestwick, E. I. Rashba,1, K. Flensberg, C. H. Stwertka, T. Taychatanapat, S. K. Watson, and C. M. Marcus,
Phys. Rev. Lett. {\bf 102}, 166802 (2009).

\bibitem{cntsodots}
B. Wunsch, Phys. Rev. B {\bf 79}, 235408 (2009); A. Secchi and M. Rontani,
Phys. Rev. B {\bf 80}, 041404(R) (2009);
M.R. Galpin, F.W. Jayatilaka, D.E. Logan, and F.B. Anders,
Phys. Rev. B {\bf 81}, 075437 (2010).

\bibitem{ando}
For reviews, see:
T. Ando, J. Phys. Soc. Jpn. {\bf 74}, 777 (2005);
J.C. Charlier, X. Blase, and S. Roche, Rev. Mod. Phys. {\bf 79}, 677 (2007).

\bibitem{gogolin}
For a textbook discussion, see A.O. Gogolin, A.A. Nersesyan, and A.M. Tsvelik, 
{\sl Bosonization and Strongly Correlated Systems}
(Cambridge University Press, Cambridge, 1998).

\bibitem{egger97}
R. Egger and A.O. Gogolin, Phys. Rev. Lett. {\bf 79}, 5082 (1997);
Eur. Phys. J. B {\bf 3}, 281 (1998); C. Kane, L. Balents, and
M.P.A. Fisher, Phys. Rev. Lett. {\bf 79}, 5086 (1997). 

\bibitem{bockrath}
M. Bockrath \textit{et al.},
Nature {\bf 397}, 598 (1999);
Z. Yao, H.W.C. Postma, L. Balents, and C. Dekker, Nature {\bf 402}, 273 (1999);
B. Gao, A. Komnik, R. Egger, D.C. Glattli, and A. Bachtold,
Phys. Rev. Lett. {\bf 92}, 216804 (2004).

\bibitem{ishii}
H. Ishii \textit{et al.}, Nature {\bf 426}, 540 (2003).

\bibitem{spinchargeexp}
O.M. Auslaender \textit{et al.}, 
Science {\bf 308}, 88 (2005);
Y. Jompol \textit{et al.}, Science {\bf 325}, 597 (2009).

\bibitem{meden}
V. Meden and K. Sch\"onhammer, Phys. Rev. B {\bf 46}, 15753 (1992);
Phys. Rev. B {\bf 47}, 16205 (1993); J. Voit, Phys. Rev. B {\bf 47} 6740 (1993).

\bibitem{moroz}
A.V. Moroz, K.V. Samokhin, and C.H.W. Barnes, 
Phys. Rev. Lett. {\bf 84}, 4164 (2000); Phys. Rev. B {\bf 62}, 16900 (2000).

\bibitem{epl}
A. De Martino and R. Egger, Europhys. Lett. {\bf 56}, 570 (2001).

\bibitem{governale}
M. Governale and U. Z\"ulicke, Phys. Rev. B {\bf 66}, 073311 (2002).

\bibitem{hausler}
W. H\"ausler, Phys. Rev. B {\bf 70}, 115313 (2004).

\bibitem{gritsev}
V. Gritsev, G.I. Japaridze, M. Pletyukhov, and D. Baeriswyl,
Phys. Rev. Lett. {\bf 94}, 137207 (2005).

\bibitem{oleg1}
J. Sun, S. Gangadharaiah, and  O.A. Starykh,
Phys. Rev. Lett. {\bf 98}, 126408 (2007).

\bibitem{rkky}
A. Schulz, A. De Martino, P. Ingenhoven, and R. Egger, Phys. Rev. B
{\bf 79}, 205432 (2009).

\bibitem{tsvelik} A. De Martino, R. Egger, and A.M. Tsvelik,
Phys. Rev. Lett. {\bf 97}, 076402 (2006).

\bibitem{vishvesh}
W. DeGottardi, T.-C. Wei, and S. Vishveshwara, 
Phys. Rev. B {\bf 79}, 205421 (2009).

\bibitem{foot2} 
Electron-electron backscattering effects
in SWNTs are tiny\cite{egger97} and disregarded here. Moreover, we stay 
away from half-filling such that Umklapp scattering processes
also play no role.

\bibitem{kimura}
T. Kimura, K. Kuroki, and H. Aoki, Phys. Rev. B  {\bf 53}, 9572 (1996).
\end{thebibliography}
\end{document}